\documentclass[graybox]{svmult}

\usepackage{type1cm}        
\usepackage{makeidx}         
\usepackage{graphicx}        
\usepackage{multicol}        
\usepackage[bottom]{footmisc}

\usepackage{newtxtext}       %
\usepackage{newtxmath}       

\makeindex             

\newcommand{\dsp}{\displaystyle}
\newcommand{\vs}[1]{\vspace{#1 ex}}
\newcommand{\hs}[1]{\hspace{#1 em}}
\newcommand{\bfr}{\begin{flushright}}
\newcommand{\efr}{\end{flushright}}
\newcommand{\bc}{\begin{center}}
\newcommand{\ec}{\end{center}}
\newcommand{\ben}{\begin{enumerate}}
\newcommand{\een}{\end{enumerate}}

\newcommand{\be}{\begin{equation}}
\newcommand{\ee}{\end{equation}}
\newcommand{\ba}{\begin{array}}
\newcommand{\ea}{\end{array}}
\newcommand{\ct}{\cite}
\newcommand{\bit}{\bibitem}
\newcommand{\dd}[2]{\frac{\partial{#1}}{\partial{#2}}}

\newcommand{\gam}{\gamma}
\newcommand{\del}{\delta}

\newcommand{\ve}{\varepsilon}

\newcommand{\kg}{\kappa}
\newcommand{\lb}{\lambda}
\newcommand{\sg}{\sigma}

\newcommand{\vf}{\varphi}
\newcommand{\og}{\omega}
\newcommand{\Gam}{\Gamma}
\newcommand{\Del}{\Delta}

\newcommand{\Fg}{\Phi}

\newcommand{\Lb}{\Lambda}

\newcommand{\bfv}{\bf{v}}

\newcommand{\bfB}{\bf B}

\newcommand{\bfE}{\bf E}

\newcommand{\lh}{\left(}
\newcommand{\rh}{\right)}
\newcommand{\ld}{\left.}
\newcommand{\rd}{\right.}

\newcommand{\nb}{\nabla}

\newcommand{\der}{\partial}

\begin{document}

\title*{The gravity of light}
\author{Jan W.\ van Holten}
\institute{J.W.\ van Holten \at Nikhef, Science Park 105, Amsterdam NL \\
 and Lorentz Insitute, Leiden University, Leiden NL; \email{v.holten@nikhef.nl}
}
%
%
\maketitle

\abstract*{The gravitational field of an idealized plane-wave solution of the Maxwell equations 
can be described in closed form. After discussing this particular solution of the Einstein-Maxwell 
equations, the motion of neutral test particles, which are sensitive only to the gravitational background field, is analyzed. This is followed by a corresponding analysis of the dynamics of 
neutral fields in the particular Einstein-Maxwell background, considering scalars, Majorana 
spinors and abelian vector fields, respectively.}

\abstract{The gravitational field of an idealized plane-wave solution of the Maxwell equations 
can be described in closed form. After discussing this particular solution of the Einstein-Maxwell 
equations, the motion of neutral test particles, which are sensitive only to the gravitational background field, is analyzed. This is followed by a corresponding analysis of the dynamics of 
neutral fields in the particular Einstein-Maxwell background, considering scalars, Majorana 
spinors and abelian vector fields, respectively.}

\section{Light and gravity \label{s1}}

Light and gravity provide the main tools for studying the universe at large; gravity, as it
determines the interactions and paths of celestial bodies, and light as it makes them 
visible to us and enables us to unravel their properties. The theoretical descriptions of
light, as a form of electromagnetism, and gravity have much in common. The classical 
theories of electromagnetism and gravity are both local relativistic field theories; these 
fields carry physical degrees of freedom propagating energy, momentum and angular 
momentum at a finite speed $c$, commonly referred to as the speed of light, even though 
gravity, and also the color charges of subatomic particles, propagate their interactions 
at the same universal speed as well\footnote{For a discussion of the role of the universal
 constant $c$ characterizing the relations between inertial frames, see ref.\ \ct{jwvh:2021}}. 

Of course, at the microscopic level electromagnetism is more than a classical field 
theory, as quantum effects become essential to its propagation and interaction with 
matter in the form of electrons and other charged particles. A similar change in the 
way gravity behaves in this domain is expected as well, although experimental 
confirmation of these ideas has as yet remained out of reach. 

Even though the sources of gravity and electromagnetism are different, with gravity 
coupling to the local density of energy and momentum, and electromagnetism to the 
local density of electric charges and currents, the corresponding physical degrees of
freedom (classical fields in the macroscopic world) do influence each other, but in an 
asymmetric way. The present chapter is dedicated to a discussion of some aspects of 
this mutual interaction. Unless specified otherwise (when numerical estimates are 
required) units are used in which the speed of light is unity: $c = 1$.

\section{Einstein-Maxwell theory \label{s2}}

General Relativity (GR), the classical theory of gravity, states that space-time is 
endowed with a geometry, encoded in the metric $g_{\mu\nu}$, determined by 
the distribution of all combined energy- and momentum-densities. This geometry 
expresses itself in the motion of matter and light in the universe. For the interaction 
between gravity and electromagnetic fields this results in an Einstein equation 
specifying the Ricci curvature in terms of the electro-magnetic energy-momentum 
tensor: 
\be
R_{\mu\nu} - \frac{1}{2}\, g_{\mu\nu} R = - 8\pi G T_{\mu\nu}[F],
\label{vh1.1}
\ee
with the local energy-momentum density of electromagnetic fields given by 
\be
T_{\mu\nu}[F] = F_{\mu\lb} F_{\nu}^{\;\,\lb} - \frac{1}{4}\, g_{\mu\nu} F_{\kg\lb} F^{\kg\lb}.
\label{vh1.2}
\ee
At the same time the classical dynamics of the electromagnetic field is specified 
by the generalized Maxwell equations in a space-time with given dynamical metric:
\be
D_{\mu} F^{\mu\nu} = \der_{\mu} F^{\mu\nu} + \Gam_{\mu\lb}^{\;\;\;\,\mu} F^{\lb\nu}
 + \Gam_{\mu\lb}^{\;\;\;\,\nu} F^{\mu\lb} = - j^{\nu}, 
\label{vh1.3}
\ee
where $j^{\nu}$ is the electric charge-current density, and $\Gam_{\mu\lb}^{\;\;\;\,\nu}$ 
the Riemann-Christoffel connection. In the absence of charges and currents: 
$j^{\nu} = 0$, equations (\ref{vh1.1})-(\ref{vh1.3}) form a closed system describing 
gravity interacting with dynamical electromagnetic fields in otherwise empty space; 
this set-up applies in particular to the coupling of gravity with electromagnetic radiation. 

In places where the energy-momentum density of the electromagnetic field 
is small compared to the Ricci-curvature determined by external sources, such 
as the sun or compact bodies like neutron stars or black holes, one can to first 
approximation neglect the contribution of the electromagnetic fields to the 
curvature and describe the electromagnetic fields outside the external source 
regions by Maxwell equations in the gravitational background of the external 
sources. This approach is usually taken in studies of gravitational lensing, 
which offered one of the first tests of GR: observing the bending of light by the 
sun \ct{eddington:1919}; and in a more extreme case the recent observations 
of a black-hole shadow by the Event Horizon Telescope \ct{eht:2019}.

However, as equation (\ref{vh1.1}) indicates, curvature can also be induced by
electromagnetic fields themselves, even though this requires rather extreme 
electromagnetic energy densities. Indeed, according to this equation (temporarily 
reinstating the speed of light $c$) the curvature $R$ measured in $1/m^2$ 
corresponding to an energy flux $\Fg$ in $W/m^2$ is numerically of the order 
\be
\frac{R}{1/\mbox{m$^2$}} \sim\, \frac{8\pi G}{c^5}\, \Fg \simeq 
2 \times 10^{-52}\, \frac{\Fg}{1\, \mbox{W/m$^2$}}.
\label{vh1.4}
\ee
Measuring the curvature due to even intense electromagnetic radiation will 
therefore be an even more extreme challenge than the curvature due to 
the collision of very distant black holes and neutron stars \ct{ligo:2016}. In 
the following sections a more precise analysis is presented. 

\section{Plane waves \label{s3}}

An complete radiative solution of the Einstein-Maxwell equations is that of a 
plane electromagnetic wave of infinite width, which is accompanied by a parallel 
plane gravitational wave of $pp$-type \ct{brinkmann:1923}-\ct{jwvh:2018}.
With the wave propagating in the $z$-direction, it is convenient to use light-cone 
co-ordinates $u = t - z$ and $v = t + z$; the traveling plane-wave solution of the 
electromagnetic field is then expressed in terms of a transverse vector potential
\be
A_i(u) = \int_{\infty}^{\infty} \frac{dk}{2\pi} \lh a_i(k) \sin ku + b_i(k) \cos ku \rh,
\label{vh2.1}
\ee
where $i = (1,2)$ labels the directions in the transverse $x$-$y$-plane;
the corresponding electric and magnetic field strengths are given by 
\be
E_i(u) = - \ve_{ij} B_j(u) = F_{ui}(u).
\label{vh2.2}
\ee
Such solutions can take the form of wave packets of finite lengths carrying a finite 
energy flux per unit area. Specifically in the absence of external sources of curvature 
the energy density in the transverse plane is constant and the transverse geometry 
can be taken to be flat; the energy flux is then given in terms of the energy-momentum
tensor by the only non-zero component
\be
T_{uu}(u) = F_{ui} F_u^{\;\,i} = \frac{1}{2} \lh \bfE^2 + \bfB^2 \rh(u),
\label{vh2.3}
\ee
provided the metric is of the Brinkmann type
\be
ds^2 = - du dv - \Fg(u,x^i) du^2 + dx^{i\,2},
\label{vh2.4}
\ee
which is flat in the $x$-$y$-plane as required. In this Brinkmann geometry the 
only non-zero components of the Riemann curvature and Ricci tensor are
\be
R_{uiuj} = - \frac{1}{2}\, \der_i \der_j \Fg, \hs{2} 
R_{uu} = - \frac{1}{2} \lh \der_x^2 + \der_y^2 \rh \Fg.
\label{vh2.5}
\ee
The Einstein equation (\ref{vh1.1}) then reduces to a single equation linking the 
$uu$-components of the Ricci and energy-momentum tensor:
\be
\lh \der_x^2 + \der_y^2 \rh \Fg = 8 \pi G \lh \bfE^2 + \bfB^2 \rh. 
\label{vh2.6}
\ee
The general solution of this inhomogeneous linear equation for the gravitational 
potential, the metric component $\Fg(u,x^i)$, is
\be
\Fg = 2 \pi G \lh x^2 + y^2 \rh \lh \bfE^2 + \bfB^2 \rh + \Fg_0,
\label{vh2.7}
\ee
where $\Fg_0(u,x^i)$ is an arbitrary solution of the homogeneous equation
\be
\lh \der_x^2 + \der_y^2 \rh \Fg_0 = 0.
\label{vh2.8}
\ee
These solutions of the homogeneous equation represent pure gravitational waves 
of $pp$-type \ct{brinkmann:1923}: 
\be
\Fg_0 = \kg_+(u) \lh x^2 - y^2 \rh + 2 \kg_{\times}(u) xy. 
\label{vh2.9}
\ee
Equation (\ref{vh2.5}) then implies that the co-efficients $\kg_{+,\times}(u)$ represent 
the components of the corresponding Riemann tensor in the transverse plane: 
\be
R^{(0)}_{uiuj} = - \lh \ba{cc} \kg_+ & \kg_{\times} \\ \kg_{\times} & \kg_+ \ea \rh.
\label{vh2.10}
\ee
Under a rotation in the transverse plane over an angle $\vf$ they transform as 
quadrupole components: 
\be
\lh \ba{l} \kg'_+ \\ \kg'_{\times} \ea\rh = 
 \lh \ba{cc} \cos 2\vf & - \sin 2 \vf \\ \sin 2 \vf & \cos 2 \vf \ea \rh \lh \ba{l} \kg_+ \\ \kg_{\times} \ea \rh.
\label{vh2.11}
\ee
In contrast, the special solution (\ref{vh2.7}) of the inhomogenous equation proportional 
to the energy density of the electromagnetic field: $\Fg - \Fg_0 \sim x^2 + y^2$, is of 
monopole type, being invariant under rotations in the transverse plane. Thus the plane 
electromagnetic wave is accompanied by a scalar gravitational wave, on which a free 
gravitational wave of quadrupole type can be superimposed. 

\section{Motion in the background of a plane wave \label{s4}}

Classical motion of electrically neutral particles in the background of this specific 
gravitational wave is described in terms of geodesics \ct{jwvh:2011,jwvh:2018}. The 
worldline $X^{\mu}(\tau)$ of a massive particle parametrized by the proper time 
$\tau$ is restricted by the constraint
\be
\dot{U} \dot{V} - \Fg(U,X^i) \dot{U}^2 + \dot{X}^{i\,2} = 1,
\label{vh3.1}
\ee
with the overdot denoting a proper-time derivative. This constraint is one of the integrals 
of motion of the geodesic equation 
\be
\ddot{X}^{\mu} + \Gam_{\lb\nu}^{\;\;\;\mu}(X) \dot{X}^{\lb} \dot{X}^{\nu} = 0,
\label{vh3.2}
\ee
The existence of a Killing vector of the metric defined by $\der_v$ implies 
another constant of motion
\be
\dot{U} = \gam = \mbox{constant}.
\label{vh3.3}
\ee
As by definition of the laboratory velocity $v^a = dX^a/dT$ we get
\be
\frac{dU}{dT} = 1 - v_z,
\label{vh3.4}
\ee
it follows from a rewriting of the constraint (\ref{vh3.1}) that 
\be
\frac{1 - \bfv^2}{(1 - v_z)^2} + \Fg = \frac{1}{\gam^2}.
\label{vh3.5}
\ee
As $dU = \gam d\tau$, the geodesic equations in the transverse plane can be written 
alternatively as
\be
\dd{^2 X^i}{U^2} + \frac{1}{2}\, \dd{\Fg}{X^i} = 0.
\label{vh3.6}
\ee
In particular for the electromagnetic wave (\ref{vh2.7}) these equations simplify to those 
of a 2-dimensional parametric oscillator: 
\be
\dd{^2 X^i}{U^2} + 2 \pi G \lh \bfE^2 + \bfB^2 \rh X^i = 0.
\label{vh3.7}
\ee
In the special case $2 \pi G (\bfE^2 + \bfB^2) = \mu^2 =$ constant the solutions take the form 
\be
X^i(U) = X_0^i \cos \mu (U - U_0).
\label{vh3.8}
\ee
In this case the non-zero components of the Riemann tensor are 
\be
R_{uiuj} = - \mu^2 \del_{ij}.
\label{vh3.9}
\ee
Therefore the remarkable consequence is, that the geodesics oscillate in the transverse 
plane at a frequency proportional to the square root of the curvature.
Scattering of neutral test particles with a wavetrain of finite length has been 
discussed in this formalism in refs.\ \ct{jwvh:2011, jwvh:2018} and references therein.

\section{Scalar fields in a plane-wave background \label{s5}}

The gravitational field of the light wave can be probed by neutral test particles,
as discussed in the previous section, or by electrically neutral fields of scalar, vector 
or spinor type, which at the classical level are sensitive only to the non-trivial 
gravitational background. At the quantum level these fields can describe e.g.\ 
pions, photons or neutrinos. As a first example we discuss a massive real scalar 
field $S$, with action 
\be
\ba{lll}
I[S] & = & \dsp{ - \frac{1}{2}\, \int d^4 x\, \sqrt{-g} \lh g^{\mu\nu}  \der_{\mu} S \der_{\nu} S 
 + m^2 S^2 \rh }\\
 & & \\
 & = & \dsp{ \int d^4 x \lh 2 \der_u S \der_v S - 2 \Fg (\der_v S)^2 - 
  \frac{1}{2} (\der_x S)^2 - \frac{1}{2} (\der_y S)^2 - \frac{m^2}{2}\, S^2 \rh, }
\ea
\label{vh4.1}
\ee
where $\nb_{\perp}$ represents the gradient in the transverse plane. The corresponding 
field equation is 
\be 
\lh 4 \der_u \der_v - 4 \Fg \der_v^2 - \der_x^2 - \der_y^2 + m^2 \rh S = 0. 
\label{vh4.2}
\ee
To solve this equation we introduce the expansion 
\be
S(u,v,x_i) = \int \frac{dqds}{2\pi}\, \kg(q,s; x_i) e^{i(qv + su)},
\label{vh4.3}
\ee
with $\kg^*(q,s;x_i) = \kg(-q,-s;x_i)$. Note that in terms if standard space-time 
co-ordinates we get 
\be
qv + su = (q+s)t + (q - s) z \equiv Et + pz \hs{1} \Leftrightarrow \hs{1}
q = \frac{1}{2} \lh E + p \rh, \hs{1} s = \frac{1}{2} \lh E - p \rh.
\label{vh4.4}
\ee
The field equation then contstrains the amplitudes in the expansion to solutions of 
\be
\lh - \der_x^2 - \der_y^2 + 4 q^2 \Fg - 4sq + m^2 \rh \kg = 0. 
\label{vh4.5}
\ee
For the special case (\ref{vh3.9}) with constant energy density $T_{uu}$ this equation 
becomes that of a 2-dimensional harmonic quantum oscillator:
\be
\lh - \der_x^2 - \der_y^2 + \og_q^2 (x^2 + y^2) - 4sq + m^2 \rh \kg = 0,
\label{vh4.6}
\ee
where $\og_q = 2 \mu |q|$. Introducing the notation $\xi_i = \sqrt{\og_q}\, x_i$
the solutions are of the form
\be
\kg(q,s;x_i) = \sum_{n_1, n_2 = 0}^{\infty} c_{n_1 n_2}(q,s) H_{n_1}(\xi_1) H_{n_2}(\xi_2)
 e^{-(\xi_1^2 + \xi_2^2)/2},
\label{vh4.7}
\ee
where the $H_n(\xi)$ are standard Hermite polynomials. It follows that the energy 
and momentum dispersion relation is quantized according to 
\be
E^2 = p^2 + m^2 + 2 (n_1 + n_2 + 1) \og_q.
\label{vh4.8}
\ee

\section{Spinor fields in a plane-wave background \label{s6}}

Our second example is a Majorana spinor field $\Psi = \Psi^c \equiv C \bar{\Psi}^T$,
where $C$ is the charge conjugation operator and $T$ denotes transposition in 
spinor space (for our conventions on the Dirac algebra including charge conjugation, 
see the appendix); such a field can describe e.g.\ neutrinos of Majorana type. 

As spinor fields are primarily defined in Minkowski space, we need to introduce the 
formalism of translating between the curved space-time manifold and the flat local
tangent space-time; this is achieved by the use of vierbein-fields $e_{\;\mu}^a$ 
such that 
\be
g_{\mu\nu} = \eta_{ab} e_{\;\mu}^a  e_{\;\nu}^b. 
\label{vh6.1}
\ee
Here $a$ labels vector components in the local Minkowski space, and $\mu$ does the 
same thing in the curved space-time manifold. Using the vierbein fields one can 
define 1-forms $E^a = e^a_{\;\mu} dx^{\mu}$, which for the metric (\ref{vh2.4}) have
the component form
\be
E^a = \lh \frac{1}{2} \lh dv + (\Fg + 1) du \rh, dx, dy, \frac{1}{2} \lh dv + (\Fg - 1) du \rh \rh.
\label{vh6.2}
\ee
Defining the inverse vierbein $e^{\mu}_{\;a}$ by 
\be
e^{\mu}_{\;a} e^a_{\;\nu} = \del_{\nu}^{\mu},
\label{vh6.3}
\ee
there is a corresponding gradient operator 
\be
\nb_a = e_{\;a}^{\mu} \der_{\mu} = 
 \lh \der_u + (1 - \Fg) \der_v, \der_x, \der_y, - \der_u + (1 + \Fg) \der_v \rh.
\label{vh6.4}
\ee
In order for the metric to be covariantly constant, the vierbein must satisfy the 
more general condition 
\be 
dE^a = \og^a_{\;b} \wedge E^b,
\label{vh6.5}
\ee
where the antisymmetric-tensor valued 1-form 
$\og^{ab} = - \og^{ba} = \og_{\mu}^{\;ab} dx^{\mu}$ defines the spin connection. 
For the special vierbein (\ref{vh6.2}) it is reduced to the form 
\be
\og^a_{\;b} = \og_{u\;b}^{\;a}\, du, \hs{2} \og_{u\;b}^{\;a} = 
 - \frac{1}{2} \lh \ba{cccc} 0 & \der_x \Fg & \der_y \Fg & 0 \\
                                                                   \der_x \Fg & 0 & 0 & - \der_x \Fg \\
                                                                   \der_y \Fg & 0 & 0 & - \der_y \Fg \\
                                                                   0 & \der_x \Fg & \der_y \Fg & 0 \ea \rh,
\label{vh6.6}
\ee
modulo an arbitrary local Lorentz transformation in tangent space. 

Defining $\og_a^{\;bc} = \og_{\mu}^{\;bc} e^{\mu}_{\;a}$, the curved-space Dirac 
operator now is 
\be
\gam \cdot D = \gam^a \lh \nb_a - \frac{1}{2}\, \og_a^{\;bc} \sg_{bc} \rh.
\label{vh6.7}
\ee
For our special metrics (\ref{vh2.4}) or vierbeins (\ref{vh6.2}) a great simplification
is, that the spin-connection term $\og_a^{\;bc} \sg_{bc}$ actually vanishes after 
contraction with $\gam^a$ \ct{jwvh:1999}; therefore the Dirac operator simplifies to 
\be
\gam \cdot D = \gam^a \nb_a = - i \lh \ba{cc} 0 & \sg_i \nb_i + \nb_0 \\
                                                                       - \sg_i \nb_i + \nb_0 & 0 \ea \rh,
\label{vh6.8}
\ee
and in the 2-component notation introduced in the appendix the Dirac equation becomes: 
\be
\ba{l}
2 \der_v \chi_2 - \lh \der_x + i \der_y \rh \chi_1 = i m \chi^*_1, \\
 \\
\lh \der_x - i \der_y \rh \chi_2 - 2 \lh \der_u - \Fg \der_v \rh \chi_1 = i m \chi^*_2.
\ea
\label{vh6.9}
\ee
The complex conjugate components $\chi^*$ can be eliminated by applying the 
complex conjugate Dirac equation to get 
\be
\ba{l}
\lh 4 \der_u \der_v - 4 \Fg \der_v^2 - \der_x^2 - \der_y^2 + m^2 \rh \chi_1 = 0, \\
 \\
\lh 4 \der_u \der_v - 4 \Fg \der_v^2 - \der_x^2 - \der_y^2 + m^2 \rh \chi_2 = 
 2 \lh \der_x + i \der_y \rh \Fg\, \der_v \chi_1.
\ea
\label{vh6.10}
\ee
Therefore the general solution for $\chi_1$ is fully analoguous to that for the 
scalar field $S$, with the same spectrum of energy and momentum states; in 
contrast, the general solution for $\chi_2$ consists of a special solution, defined 
in terms of the solution for $\chi_1$ by the right-hand side of the second equation 
(\ref{vh6.10}), plus an arbitrary solution of the homogeneous free Klein-Gordon 
equation, as for $\chi_1$. Therefore all solutions are found to have a structure 
similar to the scalar field, except that any non-trivial solution $\chi_1$ is 
accompanied by a special dependend solution for $\chi_2$ constructed from 
$\chi_1$ by  
\be
\left[ - \lh \der_x^2 + \der_y^2 \rh + m^2 \right] \chi_2 = 
 2(\der_x + i \der_y) \lh \der_u - \Fg \der_v \rh \chi_1 - 2 im \lh \der_u - \Fg \der_v \rh \chi_1^*.
\label{vh6.11}
\ee

\section{Massless abelian vector fields in a plane-wave background \label{s7}} 

Finally we describe the propagation of a massless abelian vector field in the 
plane-wave gravitational background. The Maxwell-action takes the form 
\be
\ba{lll}
\dsp{ I[a] = \int dudvdxdy }& & \left[ (\der_u a_v - \der_v a_u)^2 + 
   (\der_u a_i - \der_i a_u)(\der_v a_i - \der_i a_v) \rd \\
 & & \\
 & & \hs{1} \ld -\, \Fg (\der_v a_i - \der_i a_v)^2 - \frac{1}{8} \lh \der_i a_j - \der_j a_i \rh^2 \right]. 
\ea
\label{vh7.1}
\ee
The resulting field equations are 
\be
\ba{l}
\dsp{ 4 \der_u \der_v a_v - \Del_{\perp} a_v - 
    2 \der_v \lh \der_u a_v + \der_v a_u - \frac{1}{2}\, \der_i a_i \rh = 0, }\\
 \\
\dsp{ 4 \der_u \der_v a_u - \Del_{\perp} a_u  - 2 \der_u \lh  \der_u a_v + \der_v a_u 
 - \frac{1}{2}\, \der_i a_i \rh + 2 \der_i \left[ \Fg \lh \der_i a_v - \der_v a_i \rh \right] = 0, }\\
 \\
\dsp{ - 2 \der_u \der_v a_i + \frac{1}{2}\, \Del_{\perp} a_i + \der_i \lh \der_u a_v + \der_v a_u  
 - \frac{1}{2}\, \der_j a_j \rh - 2 \der_v \left[ \Fg \lh \der_i a_v - \der_v a_i \rh \right] = 0. }
\ea   
\label{vh7.2}
\ee
Gauge transformations $a'_{\mu} = a_{\mu} + \der_{\mu} \Lb$ can be used to simplify 
these equations. First note that we can take
\be
\der_u a'_v + \der_v a'_u - \frac{1}{2}\, \der_i a'_i = 0,
\label{vh7.3}
\ee
by taking $\Lb$ as the solution of
\be
\lh - 4 \der_u \der_v + \der_x^2 + \der_y^2 \rh \Lb = 
2 \lh \der_u a_v + \der_v a_u \rh - \der_i a_i.
\label{vh7.4}
\ee
The remaining field equations are 
\be
\ba{l}
\dsp{ 4 \der_u \der_v a'_v - \lh \der_x^2 + \der_y^2 \rh a'_v = 0, }\\
 \\
\dsp{ 4 \der_u \der_v a'_u - \lh \der_x^2 + \der_y^2 \rh a'_u + 
 2 \der_i \left[ \Fg \lh \der_i a'_v - \der_v a'_i \rh \right] = 0, }\\
 \\
\dsp{ 4 \der_u \der_v a'_i - \lh \der_x^2 + \der_y^2 \rh a'_i 
  + 4 \der_v \left[ \Fg \lh \der_i a'_v - \der_v a'_i \rh \right] = 0. }
\ea   
\label{vh7.5}
\ee
Next we can still make a residual gauge transformation to eliminate $a'_v$
by taking $\Lb'$ restricted by 
\be
\lh - 4 \der_u \der_v + \der_x^2 + \der_y^2 \rh \Lb' = 0, \hs{2} 
a^{\prime\prime}_v = a'_v + \der_v \Lb' = 0.
\label{vh7.4.1}
\ee
Then the gauge contraint (\ref{vh7.3}) reduces to 
\be
\der_v a_u^{\prime\prime} = \frac{1}{2}\, \der_i a_i^{\prime\prime}.
\label{vh7.7}
\ee
Therefore we are left with 
\be
\ba{l}
\dsp{ 4 \der_u \der_v a^{\prime\prime}_u - 4 \Fg\, \der_v^2 a_u^{\prime\prime} 
 - \lh \der_x^2 + \der_y^2 \rh a^{\prime\prime}_u 
 = 2 \der_i \Fg\, \der_v a^{\prime\prime}_i, }\\
 \\
\dsp{ 4 \der_u \der_v a^{\prime\prime}_i - 4 \Fg\,  \der^2_v a^{\prime\prime}_i  
  - \lh \der_x^2 + \der_y^2 \rh a^{\prime\prime}_i= 0, }
\ea
\label{vh7.6}
\ee
This set of equations looks similar to that of the Majorana-Dirac equations in the 
previous section: the transverse components $a_i^{\prime\prime}$ are solutions 
of the scalar Klein-Gordon equation, accompanied by a special fixed solution
$\bar{a}_u^{\prime\prime}$:
\be
- (\der_x^2 + \der_y^2)\, \bar{a}_u^{\prime\prime} = 
- 2 \der_i \left[ \lh \der_u - \Fg \der_v \rh a_i^{\prime\prime} \right]. 
\label{vh7.7}
\ee
But in contrast to the Majorana-Dirac case there is no independent dynamical solution 
for $a_u^{\prime\prime}$, as for vanishing $a_i^{\prime\prime} = 0$ the homogeneous 
equation for $a_u^{\prime\prime}$ implies its vanishing as well: 
\be
\der_v a_u^{\prime\prime} = 0 \hs{1} \mbox{and} \hs{1} 
 \lh \der_x^2 + \der_y^2 \rh a_u^{\prime\prime} = 0. 
\label{vh7.8}
\ee
These conditions do not allow normalizable solutions for $a_u^{\prime\prime}$; in 
fact, for $a_i^{\prime\prime} = 0$ the longitudinal component $a_u^{\prime\prime}$ 
can be gauged away by a third residual gauge transformation with a gauge function
 $\Lb^{\prime\prime}$ satisfying the constraints
\be
a_u^{\prime\prime\prime} = a_u^{\prime\prime} + \der_u \Lb^{\prime\prime} = 0, \hs{1} 
\der_v \Lb^{\prime\prime} = 0, \hs{1} (\der_x^2 + \der_y^2)\, \Lb^{\prime\prime} = 0. 
\label{vh7.9}
\ee
Therefore the transverse components are the only dynamical ones, taking the 
same form as solutions of the massless scalar wave equation, with the same 
spectrum of energy and momentum, whilst 
$a_u^{\prime\prime} = \bar{a}_u^{\prime\prime}$ is a dependend field fixed 
entirely in terms of the transverse components by equation (\ref{vh7.7}). 

\section{Conclusions \label{s8}}

In summary, in the above it has been shown that the equations of motion of 
neutral test particles, and the field equations of neutral scalar fields, 
Majorana-Dirac spinor fields and abelian vector fields can all be solved 
in the background of gravitational $pp$-waves such as those accompanying 
infinite plane electromagnetic waves. For energy-momentum density of the 
source field constant in time, such as that of a circularly polarized plane light 
wave, a distinct signature is, that test particles oscillate in the transverse plane 
of the wave, whilst the spectrum of transverse momentum of the fields becomes 
discrete. Because of the small curvature to be expected from such waves, 
these effects will be difficult to observe; moreover, in realistic conditions 
beams of electromagnetic wave will be of finite width, introducing modifications 
to the above conclusions which still have to be considered. Nevertheless, 
as a matter of principle the scattering of neutral particles by beams of 
electromagnetic waves will be another test in establishing the universality 
and dynamics of gravitational interactions.

\section*{Appendix: Spinors and the Dirac algebra \label{a1}}
\addcontentsline{toc}{section}{Appendix}

Spinor fields in curved space-time are most easily described in the tangent 
Minkowski space, using the vierbein formulation to translate the results to the 
curved space-time manifold. We use the flat-space representation of the Dirac 
algebra in which $\gam_5$ is diagonal: 
\be
\gam_0 = i \lh \ba{cc} 0 & 1 \\
                                  1 & 0 \ea \rh, \hs{1}
\gam_i = \lh \ba{cc} 0 & - i \sg_i \\
                                i \sg_i & 0 \ea \rh, \hs{1}
\gam_5 = \lh \ba{cc} 1 & 0 \\
                                 0 & -1 \ea \rh, 
\label{a1.1}
\ee
such that for $a = (0,1,2,3)$
\be
\left\{ \gam_a, \gam_b \right\} = 2 \eta_{ab} {\bf 1}, \hs{2} \gam_5^2 = {\bf 1}, \hs{2} 
\left\{ \gam_5, \gam_a \right\} = 0.
\label{a1.2}
\ee
The generators of the Lorentz transformations on spinors are defined by 
\be
\sg_{ab} = \frac{1}{4} \left[ \gam_a, \gam_b \right], 
\label{a1.3}
\ee
with commutation relations 
\be
\left[ \sg_{ab}, \sg_{cd} \right] = \eta_{ad} \sg_{bc} - \eta_{ac} \sg_{bd} - \eta_{bd} \sg_{ac} 
 + \eta_{bc} \sg_{ad}.
\label{a1.4}
\ee
Hermitean conjugation is achieved by 
\be
\gam_a^{\dagger} = \gam_0 \gam_a \gam_0.
\label{a1.5}
\ee
The charge conjugation operator $C$ is defined by 
\be
C = C^{\dagger} = C^{-1} = - C^T = \gam_2 \gam_0 =  \lh \ba{cc} \sg_2 & 0 \\
                                                                                           0 & - \sg_2 \ea \rh,
\label{a1.6}
\ee
such that 
\be
C^{-1} \gam_a C = - \gam_a^T.
\label{a1.7}
\ee
If the spinor $\Psi$ is a solution of the Dirac equation in Minkowski space
\be
\lh \gam \cdot \der + m \rh \Psi = 0,
\label{a1.8}
\ee
then this is also true for the charge-conjugate: 
\be
\Psi^c = C \bar{\Psi}^T = - \gam_2 \Psi^* \hs{1} \Rightarrow \hs{1} 
\lh \gam \cdot \der + m \rh \Psi^c = 0.
\label{a1.9}
\ee
The Majorana constraint $\Psi^c = \Psi$ reduces the number of independent 
spinor components from 4 to 2 complex ones. This makes it easy to work 
in terms of 2-component spinors $(\chi, \eta)$ which are eigenspinors of 
$\gam_5$, by the decomposition 
\be
\Psi = \Psi^c = \left[ \ba{c} \chi \\ \eta \ea \right] \hs{2} \eta = - i \sg_2 \chi^*.
\label{a1.10}
\ee
\vs{3} 

\begin{acknowledgement}
Discussions with Ernst Traanberg of the Lorentz Institute in Leiden on solving  
field equations in the plane-wave background are gratefully acknowledged. 
\end{acknowledgement}

\end{document}